# *Partial traces in decoherence and in interpretation: What do reduced states refer to?*

Sebastian Fortin – Olimpia Lombardi

*CONICET- Facultad de Ciencias Exactas y Naturales, Universidad de Buenos Aires*

## *Abstract*

The interpretation of the concept of reduced state is a subtle issue that has relevant consequences when the task is the interpretation of quantum mechanics itself. The aim of this paper is to argue that reduced states are not the quantum states of subsystems in the same sense as quantum states are states of the whole composite system. After clearly stating the problem, our argument is developed in three stages. First, we consider the phenomenon of environment-induced decoherence as an example of the case in which the subsystems interact with each other; we show that decoherence does not solve the measurement problem precisely because the reduced state of the measuring apparatus is not its quantum state. Second, the non-interacting case is illustrated in the context of no-collapse interpretations, in which we show that certain well-known experimental results cannot be accounted for due to the fact that the reduced states of the measured system and the measuring apparatus are conceived as their quantum states. Finally, we prove that reduced states are a kind of coarse-grained states, and for this reason they cancel the correlations of the subsystem with other subsystems with which it interacts or is entangled.

## *1. Introduction*

In physics it is common to talk about composite quantum systems and their subsystems. In general, given two systems $S_1$ and $S_2$, represented in the Hilbert spaces $H_1$ and $H_2$ respectively, the system $S = S_1 + S_2$ represented in $H = H_1 \otimes H_2$ is said to be composite: $S_1$ and $S_2$ are considered as subsystems of $S$. The relationship between the composite system and its subsystems is usually defined independently from whether $S_1$ and $S_2$ interact with each other or not, and from whether the state of $S$ is entangled or not: in any case, the common practice is to conceive the quantum states of the subsystems $S_1$ and $S_2$ as their reduced states, computed on the state of the whole system $S$ by tracing over the degrees of freedom of $S_2$ and $S_1$ respectively. This is a basic assumption in the theory of environment-induced decoherence (Zurek 1981, 1982, 1991, 1993, 1994; for a review, see

Zurek 2003), which aims at explaining the emergence of classicality from the quantum realm in terms of the fast diagonalization of the reduced state of a system in interaction with its environment. Reduced states also play a central role in certain no-collapse interpretations of quantum mechanics, as the modal interpretations (see, *e.g.* Kochen 1985, Dieks 1988, 1989, 1998, 1994a, 1994b, Vermaas & Dieks 1995, Lombardi & Castagnino 2008, Ardenghi, Castagnino & Lombardi 2009, Lombardi, Castagnino & Ardenghi 2010, Ardenghi & Lombardi 2011; for a review, see Lombardi & Dieks 2012), in which it is supposed that the quantum states of the measured system and of the measuring apparatus after the measurement interaction are represented by operators resulting from partial traces.

The presence of reduced states is certainly pervasive in practice, and physicists perfectly know how to use them. Nevertheless, the interpretation of the concept of reduced state is a subtle issue that has relevant consequences when the task is the interpretation of quantum mechanics itself. In general, a reduced state is simply considered the quantum state of an open system, since it allows us to compute the expectation values of all the observables of that system, exactly like the state of the whole system with respect to all its observables. The aim of this paper is to reject that common assumption by arguing that reduced states are actually not the quantum states of subsystems in the same sense as quantum states are states of the whole composite system. After stating the general problem of interpreting the state of a mechanical theory, our argument will be developed in three stages. First, we will consider the phenomenon of environment-induced decoherence as an example of the interacting case, in order to explain that it does not solve the measurement problem precisely because the reduced state of the measuring apparatus −modeled as an open system in continuous interaction with its environment− is not its quantum state. Second, the non-interacting case will be illustrated in the context of no-collapse interpretations, in which we will show that certain well-known experimental results cannot be accounted for due to the fact that the reduced states of the measured system and the measuring apparatus are conceived as their quantum states. Finally, we will argue that reduced states are a kind of coarse-grained states, and for this reason they cancel the correlations of the subsystem with other subsystems with which it interacts or is entangled. The moral of the present work will be that the difference between quantum states and reduced states cannot be disregarded, in particular when foundational issues about the theory are considered: an interpretation of quantum mechanics should not endow quantum states and reduced states with the same meaning if conceptual puzzles are to be avoided.

## *2. Stating the problem*

A physical theory, in particular, a mechanics, is characterized by certain elementary assumptions –postulates– that supply the starting point of any development of the theory. Among these assumptions, two are unavoidable: the definition of the state and the dynamical law that rules the time evolution of that state. For instance, in classical mechanics the state is given by the position and the momentum of the body –and other degrees of freedom if it is not a point particle– and the dynamical law is Newton's Second Law. This departing point is unavoidable from the conceptual viewpoint, since an essential part of the interpretation of the theory under consideration consists in elucidating the meaning of the corresponding state.

In order to illustrate what was said above, let us consider Gibbsian statistical mechanics: in this case, the state is given by the density function $\rho(x,t)$, defined on a phase space $\Gamma$, which evolves according to the Liouville equation. During the last decades many discussions focused on the use of ensembles in the definition of the density function, the connection of the density function with the different interpretations of probability, etc. (see, *e.g.*, Frigg 2007, Uffink 2007). One of the central problems in this theoretical context is how to explain the approach to equilibrium by means of the time evolution of the density function. In fact, as it is well known, the evolution of $\rho(x,t)$ has no limit for $t \to \infty$: the volume of the support of $\rho(x,t)$ in $\Gamma$ is time-invariant as a consequence of the Liouville theorem. When classical statistical mechanics is formulated in the formalism of Koopman (1931) –its Hilbert space formulation–, that result can be expressed by saying that the vector state $\rho$ evolves unitarily according to $\rho(t) = U_t \rho_0 = e^{-iGt} \rho_0$, where $\rho_0$ is the initial state and $G$ is the generator of the evolution, which represents the energy of the system: since $U_t$ is a unitary operator, it does not change the angle of separation (the inner product) or the distance (the square modulus of the difference) between vectors representing two different states. In other words, due to its unitary evolution, $\rho(t)$ has no limit for $t \to \infty$. Therefore, the problem of irreversibility in classical statistical mechanics turns out to be how to account for a non-unitary irreversible approach to equilibrium in systems ruled by a unitary evolution law (see Lombardi 2003, Frigg 2007).

The standard answer in the context of the Gibbsian approach consists in relying on coarse-graining: the phase space $\Gamma$ is partitioned into cells $C_i$ of a same volume $c_i$ in $\Gamma$, and a coarse-grained distribution $\rho_{cg}$ is defined such that, for each $C_i$,

$$\begin{cases} \text{if } \forall x \in C_i, \rho(x) = 0 \to \rho_{cg} = 0 \\ \text{if } \exists x \in C_i, \rho(x) \neq 0 \to \rho_{cg} = 1/c_i \int_{C_i} \rho(x) \, dx \neq 0 \end{cases} \quad (1)$$

It is quite clear that, since the coarse-grained state $\rho_{cg}$ is not the original statistical state, it is not governed by the dynamical law of classical statistical mechanics and, as a consequence, its evolution is not constrained by the Liouville theorem: $\rho_{cg}(t)$ may approach a definite limit for $t \to \infty$. This is what effectively happens when the system has a sufficiently high degree of instability: if the system is mixing, it can be proved that the coarse-grained state approaches an equilibrium state, $\rho_{cg}(t) \to \rho_{cg(eq)}$ (see, *e.g.*, Berkovitz, Frigg & Kronz 2006).

Of course, discussions also focus on the meaning of the coarse-grained state: Does it supply the objective description of an emergent process? Or does it only describe our ignorance about the real underlying evolution? These questions express the different interpretations of irreversibility, in the context of a debate that is still as alive as in the birth of statistical mechanics (see, *e.g.*, Nicolis & Prigogine 1989, Lebowitz 1993, 1994a, 1994b, Driebe 1994, Bricmont 1995, Earman & Rédei 1996, Callender 1999). Nevertheless, despite the heated discussions and the deep disagreements about interpretation, nobody ignores the difference between the statistical state, which evolves unitarily according to the dynamical postulate of the theory, and the coarse-grained state, which may tend to a final stable state. Even those authors who take a heterodox position by claiming the priority of the coarse-grained state regarding objectivity, are clear that such a move requires a reformulation of the dynamical law of the theory (see, *e.g.*, Misra, Prigogine & Courbage 1979, Nicolis & Prigogine 1989). In other words, independently of the particular disagreements, nobody attempts to endow the statistical state and the coarse-grained state with the same interpretation.

What happens in quantum mechanics? As it is well known, the hottest conceptual discussions about the theory are those referred to the interpretation of the quantum state. Does it describe a physical field on physical but empirically inaccessible degrees of freedom? Or does it represent our degrees of knowledge? Or does it embody certain objective propensities to actualization? This kind of questions has led to the wide spectrum of interpretations of the theory, which continues to grow. In particular, the question of the meaning of the quantum state is a topic that attracts the interest of researchers up to the present (see, *e.g.* Ney & Albert 2013, in particular, Maudlin's article "The nature of the quantum state" in that volume). Given the open-question status of this issue, it deserves to be still discussed in detail.

Paradoxically, despite the severity of the interpretive problem, in certain contexts the distinction between the different kinds of states appearing in the quantum discourse is not sufficiently emphasized. In the standard presentations of the theory, the quantum state is represented by an operator $\rho \in H \otimes H$, where $H$ is a Hilbert space (pure states represented by vectors $|\psi\rangle \in H$ can be considered as the particular case $\rho = |\psi\rangle\langle\psi|$), and its unitary time-evolution is ruled by the von Neumann equation (or, in the particular case of pure states, by the Schrödinger equation).[1] However, in recent times this presentation has been downplayed by the irruption of the reduced states. For instance, it is said that the dynamical postulate of quantum mechanics only applies to closed systems, whereas reduced operators actually represent quantum states of open systems.[2] Nevertheless, we are not informed about the evolution *law* for reduced states; in fact, the evolution of open systems always depends, in the final analysis, on the unitary evolution of the whole closed system of which the open system is a part. It is also admitted that, whereas the states of closed systems embody quantum correlations, reduced states may cancel those correlations and, as a consequence, cannot be used for computations in certain cases.[3] However, in spite of this central difference, the states of closed and open systems are treated on equal footing. In turn, the coarse-grained nature of the reduced states is acknowledged by some authors −although without explicitly demonstrating this claim− (see, *e.g.*, Anastopoulos 2002). Nevertheless, the change in the concept of reduced state resulting from this move is not noticed: from being originally conceived as the quantum state of the *open* system, it turns out to be the coarse-grained state of the *closed* system.

In summary, reduced states are coarse-grained states that cancel correlations and evolve non-unitarily. All these peculiar features are absent from those quantum states traditionally referred to by the postulates of the theory. Therefore, disregarding these differences is at least curious: it is analogous to identify the state represented by the density function $\rho(x,t)$ with the coarse-grained state represented by $\rho_{cg}(x,t)$ in the interpretation of classical statistical mechanics; as stressed above, this is an identification that nobody establishes. The fact that physicists perfectly know how to use reduced states in quantum mechanics is not an excuse to refrain from discussing this foundational issue. Our aim in the following paragraphs is precisely to analyze the meaning of the reduced states, in order to alert regarding their uncritical use in the interpretation context.

---

[1] Here we will not discuss the question of the supposed priority of pure states over mixed states: following certain presentations of the theory (*e.g.* Ballentine 1998), we will take the generic stance of considering state operators as representing quantum states, and pure states as a particular case.

[2] Wayne Myrvold stressed this point in a comment to our talk at the *14th Congress of Logic, Methodology and Philosophy of Science* (Nancy, July 19-26, 2011). We will come back to this point in the next section.

[3] We thank one of the anonymous referees for stressing this point.

## *3. Reduced states in decoherence*

### 3.a) Environment-induced decoherence

The environment-induced decoherence program was born in the seventies, when the measurement problem began to be addressed from an open-system perspective: according to that view, macroscopic systems such as measurement devices are never closed, but interact significantly with their environments (see Zeh 1970, 1973). On the basis of those previous contributions, the theory of decoherence was systematized and developed by Zurek and his collaborators in a great number of works. According to Zurek (1982, 1991, 1993, 1994), decoherence is a process resulting from the interaction between a quantum system and its environment; this process singles out a preferred set of states, usually called "pointer basis", that determines which observables receive definite values. Therefore, the first step is to split the universe into the degrees of freedom which are of direct interest to the observer, "the system of interest", and the remaining degrees of freedom usually referred to as "the environment". The environment can be external, such as particles of air or photons scattered off the system, or internal, such as collections of phonons or other internal excitations.

As it is well known, the literature on decoherence usually considers a small system $S$ in interaction with a large environment $E$: the whole composite system $S+E$ is in a state $|\psi_{SE}(t)\rangle$ or $\rho_{SE}(t) = |\psi_{SE}(t)\rangle\langle\psi_{SE}(t)|$, where the off-diagonal terms represent the quantum correlations that preclude classicality. Of course, the quantum correlations do not vanish throughout the unitary time-evolution of the state of $S+E$. But, according to Zurek, at each time *the quantum state of the system $S$ is given by the reduced density operator $\rho_S^r(t)$*, obtained from the total density operator $\rho_{SE}(t)$ by tracing over the environmental degrees of freedom:

$$\rho_S^r(t) = Tr_{(E)}\rho_{SE}(t) \qquad (2)$$

In the phenomenon of decoherence, as time passes the composite system $S+E$ evolves under the influence of the interaction Hamiltonian $H_{SE}$ in such a way that, after an extremely short decoherence time, the reduced state $\rho_S^r(t)$ of $S$ settles down in a time-independent state $\rho_{S(dec)}^r$, where the off-diagonal terms have vanished.

According to Zurek, the incessant "monitoring" of some observables by the environment leads to the "degradation" of pure states into mixtures. Zurek conceives the *reduced state* of the decohering system $S$ as its quantum state: according to him, $\rho_{S(dec)}^r$ denotes a *quantum mixture* that contains only the terms corresponding to classical correlations. It is precisely for this reason that decoherence would offer the clue for solving the quantum measurement problem and for explaining the quantum-to-classical transition of the system $S$. In a certain sense, decoherence would explain collapse since "*quantum entanglement will be converted into an effectively classical correlation as a result of the interaction*" between the system and its environment (Paz & Zurek 2002, p. 90).

As some point out (Leggett 1987, Bub 1997), the theory of environment-induced decoherence has become the "new orthodoxy" in the quantum physicists community: many authors consider that decoherence finally supplies the right answer to the measurement problem (see, *e.g*., Auletta 2000, Anderson 2001). However, not all are so enthusiastic: the account of measurement given by the decoherence theorists has been severely criticized (see, *e.g*., Healey 1995, Bacciagaluppi 2008). For instance, Adler claims that the diagonalized reduced state $\rho^r_{S(dec)}$ does not allow us to say that the state of the system $S$ is in one of the eigenstates $|r_i\rangle$ of the observable $R$, and he concludes: "*I do not believe that either detailed theoretical calculations or recent experimental results show that decoherence has resolved the difficulties associated with quantum measurement theory*" (Adler 2003, p. 135). According to Bub (1997), if the eigenstate-eigenvalue link is accepted, the problem is even worst: the reduced state is not only unable to explain the occurrence of only one of the eigenvalues $r_i$ of $R$, but it is also inconsistent with that occurrence, since the state of the composite system $S + E$ is always the entangled state $|\psi_{SE}(t)\rangle$. These and similar arguments have led even some contributors to the decoherence program to express their skepticism about the relevance of decoherence to the solution of the measurement problem; as Joos (2000, p. 14) says: "*Does decoherence solves the measurement problem? Clearly not*." Here we will argue that a way of understanding this last claim is to notice that, by contrast to Zurek's strategy, the reduced state of the decohering system $S$ must not be conceived as its quantum state.

### 3.b) The definition of the reduced state

The interpretation of a reduced operator as representing the quantum state of a component system of a composite system is a usual practice in physics. As Dieks asserts in the case of measurement: "*The projection operator $|\psi\rangle\langle\psi|$ is an observable of the total system [...]. But we are really interested in the individual properties of device and object taken by themselves. Therefore, we need substitutes for $|\psi\rangle\langle\psi|$ that represent the states of these individual systems. In the context of standard quantum mechanics such operators are readily available, namely the density operators for the partial systems*" (Dieks 2007, p. 298); "*we took [the reduced operator] $W_1$ for the state of system 1. This is* *standard practice in quantum mechanics; however, the usual justification relies on the probabilistic interpretation of the theory and the Born rule*" (Dieks 2007, p. 299). These quotes suggest that the reduced state $W_1$ is the state of the open system in the same sense as $|\psi\rangle$ is the state of the total closed system. This is precisely the idea that we want to challenge.

Mathematically, the reduced state $\rho_1^r$ of a system $S_1$, which is a component of a larger system $S = S_1 + S_2$ represented in the Hilbert space $H = H_1 \otimes H_2$, is a density operator since it satisfies the normalization condition ($Tr\rho_1^r = 1$), the Hermitian condition ($\rho_1^{r\dagger} = \rho_1^r$) and the non-negativeness condition ($\langle\varphi|\rho_1^r|\varphi\rangle \geq 0$ for all $|\varphi\rangle \in H_1$). Nevertheless, it denotes an improper mixture (d'Espagnat 1976): although proper and improper mixtures are represented by the same kind of mathematical object −a density operator−, they refer to *physically different concepts*. In fact, if $\{|a_m\rangle\}$ is a basis of $H_1$ and $\{|b_n\rangle\}$ is a basis of $H_2$, the reduced states of $S_1$ and $S_2$ are obtained as

$$\rho_1^r = Tr_{(2)}\rho = \sum_n \langle b_n|\rho|b_n\rangle \tag{3}$$

$$\rho_2^r = Tr_{(1)}\rho = \sum_m \langle a_m|\rho|a_m\rangle \tag{4}$$

and they are sufficient to compute the expectation value of any observable that belongs exclusively to $S_1$ or $S_2$, respectively. However, these two density operators are not sufficient, in general, to determine the state $\rho$ of the composite system $S$, since they provide no information about the correlations between $S_1$ and $S_2$. Thus, if we could only make measurements on $S_1$ and could not make any on $S_2$, then we would not be able to differentiate the improper mixture denoted by the reduced density operator $\rho_1^r$ from the analog proper mixture denoted by the density operator $\rho_1$. But, as d'Espagnat stresses (1995), there is no theoretical reason that prevents us from having access to, at least, some of the traced over degrees of freedom, and this access would permit us to show that the proper mixture and the improper mixture are, in principle, testably different. In the context of decoherence, the difference is also pointed out by Zeh, one of the founding fathers of the theory: "*The conceptually important difference between true and apparent ensembles was clearly pointed out by Bernard d'Espagnat (1976) when he distinguished between proper and improper mixtures. In the case of virtual (reversible) decoherence, this difference can even be observed as recoherence (a relocalization of the superposition, that would be impossible for a proper mixture)*" (Zeh 2005, p. 2).

However, this difference between proper and improper mixtures has been considered an insufficient reason for depriving $\rho_1^r$ of the status of being the quantum state of the open system. In fact, some authors insist that improper mixtures are the quantum states of the components of a composite system, even though the quantum state of the composite system is not uniquely defined by the quantum states of its components: "[W]*hile the states of subsystems of a composite system are uniquely determined by the state of the composite system via the reduced density operator, the state of a composite system is not, in general, uniquely determined by the states of its subsystems.*" (Bub 1997, p. 265; see also Hughes 1989). But this position does not consider the difference regarding time-evolutions: whereas quantum states evolve according the Schrödinger (or the von Neumann) equation and, therefore, always follow unitary evolutions, improper mixtures in general evolve non-unitarily. This amounts to admit that, given the basic postulates of quantum mechanics, there are certain states that are not ruled by the dynamical postulate of the mechanical theory.

In this insistence on blurring the difference between proper and improper mixtures, it could still be argued that the non-unitary evolution of the reduced states does not imply the violation of one of the principles of the theory, to the extent that the dynamical postulate of quantum mechanics only applies to *closed* systems, but says nothing about the time-evolution of open systems. According to this viewpoint, reduced operators actually represent quantum states of *open* quantum systems, like those involved in measurement, whose time dependence is ruled by non-unitary evolution laws. This amounts to claim that quantum mechanics is a mechanical theory with two different kinds of states, each one with its own evolution law. This would be a novelty in physics, since the only mechanical theory without a unified state description was Aristotle's theory, with the difference between the sublunar and the supralunar world, each one with its own dynamical law: the "law" of the natural places on earth and the "law" of the uniform circular motion in heaven.

A different strategy is to claim that the true states of quantum mechanics are the reduced states, and the unitarily evolving states are a limiting case of those generic states. This is the strategy adopted by Cartwright (1983), who proposed to replace the Schrödinger (or the von Neumann) equation by the generalized master equation governing the time-evolution of the reduced state as the dynamical postulate of quantum mechanics. However, this suggestion must face the difficulty that there is no general form of the generalized master equation: it has to be derived case by case depending on the specific features of each particular situation, and always on the basis of assuming the unitary time-evolution of the closed system. Furthermore, the generalized master equation can be viewed as a convenient tool without independent theoretical meaning. In fact, as Paz and Zurek (2000, p. 25) admit, "*In principle, the evolution equation for* [the reduced state] $\rho$ *could be obtained* *by solving Schrödinger (or von Neumann) equation for the total density matrix and then taking the trace. However, this task can be analytically completed in very few cases, and the study of the evolution of the reduced density matrix should be done by using some approximations.*" In other words, in principle we could dispense with the generalized master equation by considering the unitary evolution of the whole closed system and taking the partial trace at each particular time to obtain the reduced state for that time. This means that the use of the generalized master equation for the reduced state is only due to practical limitations and always depends on the unitary evolution of the state of the closed system. As a consequence, Cartwright's proposal of conceiving the reduced state and the generalized master equation as the new pair state-dynamical law for quantum mechanics is not well founded.

The insistence on considering the reduced state (improper mixture) as the mixed quantum state of the open system seems to forget the definition of the concept of reduced state. The reduced state $\rho_1^r$ of $S_1$ is defined as the density operator by means of which the expectation values of all the observables belonging to $S_1$ can be computed. Precisely, if $H_1$ and $H_2$ are the Hilbert spaces of $S_1$ and $S_2$ respectively, $H = H_1 \otimes H_2$ is the Hilbert space of $S$, $O_1 \in H_1 \otimes H_1$ is an observable of $S_1$, $I_2$ is the identity in $H_2 \otimes H_2$, and $\rho \in H \otimes H$ is the state of $S$, then the reduced state of $S_1$ is defined as the density operator $\rho_1^r$ such that

$$\forall O = O_1 \otimes I_2 \in H \otimes H, \quad \langle O \rangle_\rho = \langle O_1 \rangle_{\rho_1^r} \tag{5}$$

On the basis of this *definition*, the reduced state can be *computed* by tracing over the degrees of freedom of $S_2$ as $\rho_1^r = Tr_{(2)} \rho$. As Schlosshauer claims in his well-known book about decoherence, this means that, strictly speaking, a reduced density operator is only "*a calculational tool*" for computing expectation values (Schlosshauer 2007, p. 48). For this reason, the author warns us "*against a misinterpretation of reduced density matrices as describing a proper mixture of states*" (2007, p. 69).

Perhaps the fact that proper and improper mixtures cannot be distinguished from a mathematical viewpoint is what leads many authors to suppose that the reduced state $\rho_1^r$ is the quantum state of the subsystem $S_1$ in the same sense as $\rho$ is the quantum state of the whole closed system $S$. However, that mathematical feature essentially depends on the formalism used to express the theory, in particular, the Hilbert space formalism; the situation might be different in other cases. For instance, it has been proved (Masillo, Scolarici & Sozzo 2009) that proper and improper mixtures are represented by different density operators in the so-called quaternionic formulation of quantum mechanics;[4] hence they can be distinguished also from a mathematical viewpoint. Moreover, this mathematical representation is compatible with their different time-evolutions as represented in the orthodox Hilbert space formalism.

### 3.c) The meaning of decoherence

Once the difference between reduced states and proper quantum states is taken into account, the usual claims surrounding decoherence can be critically assessed. For instance, Zurek uses to assert that, as a result of the interaction with the environment, "*pure states turn into mixtures and rapidly diagonalize in the einselected states*" (Zurek 2003, p. 729). But, which is the state that begins as pure and becomes mixed throughout the time-evolution? Let us suppose that Zurek is talking about the proper quantum state of the whole closed system. In this case it is true that, in the usual models, it begins as pure; however, that state never becomes mixed since it always evolves according to the unitary Schrödinger equation. If we assume, on the contrary, that the non-unitarily evolving state is the reduced state of the open system, it is true that it rapidly tends to a diagonalized mixed form; but this reduced state never begins as pure since, as it is well-known, if a closed system is in a pure state, the reduced states of its subsystems are mixtures.[5] As a consequence, although the idea that pure states become mixtures has permeated the most part of the literature about decoherence, the clear discrimination between different kinds of states allows us to identify the conceptual obscurity of this way of understanding decoherence.

If there are no pure states that become mixtures, the interpretative argument that denies the role of decoherence in the solution of the measurement problem appears in a new light. As we have seen, after an extremely short decoherence time, the reduced state $\rho_S^r(t)$ of the open system $S$ turns out to –approximately– settle down in the state $\rho_{S(dec)}^r$, where the interference terms have vanished. However, $\rho_{S(dec)}^r$ cannot be interpreted as describing the occurrence of a particular event, associated with one of the eigenstates of the apparatus pointer, with the corresponding coefficients representing a measure of our ignorance about the event that actually occurred, and the reason for this is that $\rho_{S(dec)}^r$ *does not represent the mixed quantum state of* $S$. In other words, it cannot be properly said that decoherence makes the *quantum state* of the open system to evolve to a final state that contains only the terms corresponding to classical correlations.

---

[4] The quaternionic formulation of quantum mechanics is a formalism based on quaternion fields instead of complex fields (see Adler 1995).

[5] We are grateful to one of the referees for a comment that suggested this critical remark.

Summing up, in an interpretive context, where conceptual matters must be carefully taken into account, the definition of the reduced state in terms of expectation values has always to be kept in the center of the scene. On this basis, decoherence should be described as a process through which the expectation values of all the observables belonging to the open system of interest acquire a particular final value (see Castagnino, Laura & Lombardi 2007, Castagnino, Fortin & Lombardi 2010a, Lombardi, Fortin & Castagnino 2012), without appealing to conceptually unclear evolutions that turn quantum entanglement into classical correlations and pure states into mixtures.

## *4. Reduced states in interpretation*

### 4.a) Non-interacting systems

Given a system $S = S_1 + S_2$, represented in $H = H_1 \otimes H_2$, the claim that the reduced states $\rho_1^r$ and $\rho_2^r$ are not the true quantum states of $S_1$ and $S_2$ seems to be reasonable when the components $S_1$ and $S_2$ interact with each other, that is, when the total Hamiltonian of $S$ is $H = H_1 \otimes I_2 + I_1 \otimes H_2 + H_{int}$, where $H_1, I_1 \in H_1 \otimes H_1$, $H_2, I_2 \in H_2 \otimes H_2$, and $H_{int} \neq 0 \in H \otimes H$. As we have seen, in this case both $\rho_1^r$ and $\rho_2^r$ evolve non-unitarily: this is the typical situation in decoherence. Let us now examine the non-interacting case.

Harshman and Wickramasekara (2007a, 2007b) use the term "tensor product structure" (TPS) to call any partition of the whole system $S$, represented in the Hilbert space $H$, into parts $S_1$ and $S_2$ represented in $H_1$ and $H_2$. They point out that quantum systems admit a variety of TPSs, each one of which leads to a different entanglement between the parts resulting from the partition: entanglement is TPS-dependent (see also Healey 2013). However, they also notice that there is a particular TPS that is dynamically invariant, and corresponds to the case in which there is no interaction between the subsystems $S_1$ and $S_2$. In this case, the total Hamiltonian can be decomposed as $H = H_1 \otimes I_2 + I_1 \otimes H_2$. Since $[H_1 \otimes I_2, I_1 \otimes H_2] = 0$ and $H_{int} = 0$, then,

$$exp[-iHt/\hbar] = exp[-iH_1t/\hbar] \, exp[-iH_2t/\hbar] \tag{6}$$

Therefore, if $\rho(t) = U_t \rho_0 U_{-t} = e^{-iHt/\hbar} \rho_0 e^{iHt/\hbar}$ is the quantum state of $S$, then

$$\rho_1^r(t) = Tr_{(2)} \rho(t) = Tr_{(2)} \left[ e^{-iHt/\hbar} \rho_0 e^{iHt/\hbar} \right] = e^{-iH_1t/\hbar} \left[ Tr_{(2)} \rho_0 \right] e^{iH_1t/\hbar} = e^{-iH_1t/\hbar} \rho_{10}^r e^{iH_1t/\hbar} \tag{7}$$

$$\rho_2^r(t) = Tr_{(1)} \rho(t) = Tr_{(1)} \left[ e^{-iHt/\hbar} \rho_0 e^{iHt/\hbar} \right] = e^{-iH_2t/\hbar} \left[ Tr_{(1)} \rho_0 \right] e^{iH_2t/\hbar} = e^{-iH_2t/\hbar} \rho_{20}^r e^{iH_2t/\hbar} \qquad (8)$$

This means that the subsystems $S_1$ and $S_2$ are *dynamically independent*: each one of them evolves unitarily according to the Schrödinger equation under the action of its own Hamiltonian. Moreover, in spite of the fact that $\rho(t)$, as well as $\rho_1^r(t)$ and $\rho_2^r(t)$, evolve with time, the entanglement between $S_1$ and $S_2$ is *dynamically invariant*: it cannot be expected that entanglement finally vanishes through the evolution.

The difference between the interacting and the non-interacting cases has led some authors to review the concept of subsystem in the quantum realm. For instance, in the modal-Hamiltonian interpretation (Lombardi & Castagnino 2008, Ardenghi, Castagnino & Lombardi 2009, Lombardi, Castagnino & Ardenghi 2010), when $S_1$ and $S_2$ do not follow unitary evolutions according to the dynamical law of quantum mechanics, they are not viewed as subsystems of $S$ but as mere "parts" of it. Those parts are not quantum systems because they lack independent identity: they are conceived as the result of conventional partitions of the whole quantum system $S$. Nevertheless, in the case that there is no interaction between $S_1$ and $S_2$ and their time-evolutions are governed by the Schrödinger equation, according to the modal-Hamiltonian interpretation there is no obstacle to consider them legitimate quantum systems, in particular, subsystems of the composite system $S$.

The dynamical independence of non-interacting systems might lead us to suppose that, in this particular situation, the reduced states $\rho_1^r(t)$ and $\rho_2^r(t)$ can be considered the true quantum states of $S_1$ and $S_2$ respectively. However, this assumption, implicit in several no-collapse interpretations of quantum mechanics, proves to be inadequate when certain well-known experimental results are taken into account.

**4.b) Consecutive measurements and reduced states**

In the so-called "*first kind*" measurements (Pauli 1933), the measurement interaction does not destroy the measured system. As a consequence, the question about the result of a second measurement on the same system makes sense. As it is experimentally well known, in this first kind case there are definite correlations between the outcomes of consecutive measurements.

The orthodox interpretation of quantum mechanics, which relies on the eigenstate-eigenvalue link and the hypothesis of collapse, offers a straightforward account of the outcome agreement in consecutive measurements of the same observable and of the outcome correlations in consecutive measurements of different observables (see, *e.g.*, Messiah 1961, Cohen-Tannoudji, Diu and Laloë 1977). However, since it faces difficulties with respect to other interpretive problems, different no-collapse interpretations have been proposed, according to which quantum states always follow unitary evolutions ruled by the Schrödinger (or the von Neumann) equation. These interpretations try to solve the measurement problem by selecting the set of the definite-valued observables for a system in a quantum state (for a general characterization of no-collapse interpretations, see Bub 1992, 1997, Bub & Clifton 1996). As a consequence, no-collapse interpretations usually reject the eigenstate-eigenvalue link, and allow the system to have definite-valued properties even if the system is not in an eigenstate of the observables representing those properties.

In the general von Neumann model for ideal first kind measurements, the interaction establishes a correlation between the eigenstates $|a_i\rangle$ of the observable $A$ of the measured system $S$ and the eigenstates $|r_i\rangle$ of the pointer $R$ of the measuring apparatus $M$. When the initial state of $S$ is a superposition of the $|a_i\rangle$, the state of the whole composite system $S+M$ is

$$|\psi\rangle = \sum_i c_i |a_i\rangle \otimes |r_i\rangle \tag{9}$$

If there is no collapse, the composite system preserves its state $|\psi\rangle$. Nevertheless, once the interaction ends, $S$ and $M$ turn out to be dynamically independent systems and, as stressed above, both evolve unitarily according to the Schrödinger equation. This is what may lead to conceive the reduced states $\rho_S^r = Tr_{(M)} |\psi\rangle\langle\psi|$ and $\rho_M^r = Tr_{(S)} |\psi\rangle\langle\psi|$ as the true quantum states of $S$ and $M$ respectively. This view is particularly clear in certain modal interpretations (see review in Lombardi & Dieks 2012), such as the Kochen-Dieks interpretation (Kochen 1985, Dieks 1988, 1989, 1998, 1994a, 1994b), its generalization to mixed states, given by the Vermaas-Dieks interpretation (Vermaas & Dieks 1995), and the modal-Hamiltonian interpretation (Lombardi & Castagnino 2008, Ardenghi, Castagnino & Lombardi 2009, Lombardi, Castagnino & Ardenghi 2010, Ardenghi & Lombardi 2011). The problem with this assumption is that, if taken seriously, it represents an obstacle for no-collapse interpretations to explain consecutive measurements (see detailed discussion in Ardenghi, Lombardi & Narvaja 2012).

Let us consider a first measurement performed on $S$ by a measuring apparatus $M^{(1)}$. The reduced states in this case are

$$\rho_S^{r(1)} = Tr_{(M^{(1)})} |\psi^{(1)}\rangle\langle\psi^{(1)}| = \sum_i |c_i|^2 |a_i\rangle\langle a_i| \tag{10}$$

$$\rho_M^{r(1)} = Tr_{(S)} |\psi^{(1)}\rangle\langle\psi^{(1)}| = \sum_i |c_i|^2 |r_i^{(1)}\rangle\langle r_i^{(1)}| \tag{11}$$

According to certain no-collapse interpretations, the form of $\rho_M^{r(1)}$ is what allows us to say that, although the state of the composite system $S + M^{(1)}$ does not collapse, the pointer $R^{(1)}$ of the apparatus $M^{(1)}$ indeterministically acquires a certain value $r_k^{(1)}$ with probability $|c_k|^2$. Now, let us consider a second measurement performed on $S$ by a measuring apparatus $M^{(2)}$ with a pointer $R^{(2)}$. If $\rho_S^{r(1)}$ were actually the quantum state of $S$, then the second measurement of the observable $A$ of $S$ would establish the following correlation:

$$\rho_0^{(2)} = \rho_S^{r(1)} \otimes |r_0^{(2)}\rangle\langle r_0^{(2)}| \xrightarrow{\text{interaction}} \rho^{(2)} = \sum_i |c_i|^2 |a_i\rangle\langle a_i| \otimes |r_i^{(2)}\rangle\langle r_i^{(2)}| \tag{12}$$

The reduced states in this case would be

$$\rho_S^{r(2)} = Tr_{(M^{(2)})} \rho^{(2)} = \sum_i |c_i|^2 |a_i\rangle\langle a_i| \tag{13}$$

$$\rho_M^{r(2)} = Tr_{(S)} \rho^{(2)} = \sum_i |c_i|^2 |r_i^{(2)}\rangle\langle r_i^{(2)}| \tag{14}$$

Here we should say that the pointer $R^{(2)}$ of the apparatus $M^{(2)}$ indeterministically acquires a certain value $r_l^{(2)}$ with probability $|c_l|^2$. Therefore, if we took seriously the assumption that $\rho_M^{r(1)}$ and $\rho_M^{r(2)}$ are the quantum states of the apparatuses $M^{(1)}$ and $M^{(2)}$, we would have no way of explaining the agreement between the reading $r_k^{(1)}$ of $R^{(1)}$ in the first measurement and the reading $r_l^{(2)}$ of $R^{(2)}$ in the second measurement: the partial trace has cancelled the correlations among the systems involved in the measurement process (Ardenghi, Lombardi & Narvaja 2012). This conclusion, obtained in the case of consecutive measurements of the same observable, can be easily generalized for consecutive measurements of different observables.

The professional physicist, trained in the framework of the collapse interpretation −or, sometimes, of the ensemble interpretation− immediately notices that the use of partial traces to compute correlations between the readings of correlative measurements in general leads to wrong results.[6] On this basis, he might conclude that these results are the clear demonstration that quantum measurements cannot be correctly explained without the collapse hypothesis. However, this is not the case: the correlations between the outcomes of consecutive measurements can be easily accounted for without collapse when the measuring apparatuses are taken into account and, at the same time, *partial traces are dropped* in such a way that the only legitimate quantum states are the states of the whole closed system.

---

[6] This was the reaction of Rodolfo Gambini when we explained him the basics of modal interpretations. We are grateful to him for the interesting discussion that followed.

**4.c) Consecutive measurements without collapse**

Let us consider a Stern-Gerlach experiment –paradigmatic example of first kind measurement–, where the observable $A$ to be measured is the spin $S^z$ of a particle, with eigenvectors $|z_+\rangle$ and $|z_-\rangle$, and the role of the pointer $R^{(1)}$ in the first measurement is played by the particle's momentum $P^z$ in $z$-direction, with eigenvectors $|p_0^z\rangle$, $|p_+^z\rangle$ and $|p_-^z\rangle$. Let us suppose that, initially, the particle is in a superposition of $|z_+\rangle$ and $|z_-\rangle$, and its momentum in $z$-direction is given by $|p_0^z\rangle$. The interaction with the magnetic field establishes a correlation such that

$$|\psi_0^{(1)}\rangle = (c_+|z_+\rangle + c_-|z_-\rangle) \otimes |p_0^z\rangle \longrightarrow |\psi^{(1)}\rangle = c_+|z_+\rangle \otimes |p_+^z\rangle + c_-|z_-\rangle \otimes |p_-^z\rangle \qquad (15)$$

If a second measurement is performed on the particle, now on the spin $S^x$, the role of the pointer $R^{(2)}$ in this second measurement is played by the particle's momentum $P^x$ in $x$-direction, with eigenvectors $|p_0^x\rangle$, $|p_+^x\rangle$ and $|p_-^x\rangle$. Let us suppose that, before the interaction, the second $S^x$-measuring apparatus is in its ready-to measure state $|p_0^x\rangle$:

$$|\psi_0^{(2)}\rangle = |\psi^{(1)}\rangle \otimes |p_0^x\rangle = \left(c_+|z_+\rangle \otimes |p_+^z\rangle + c_-|z_-\rangle \otimes |p_-^z\rangle\right) \otimes |p_0^x\rangle \qquad (16)$$

In order to show the correlations to be introduced by the second measurement, the states $|z_+\rangle$ and $|z_-\rangle$ must be expressed in the basis $\{|x_+\rangle, |x_-\rangle\}$:

$$|z_+\rangle = \frac{1}{\sqrt{2}}(|x_+\rangle + |x_-\rangle) \qquad |z_-\rangle = \frac{1}{\sqrt{2}}(|x_+\rangle - |x_-\rangle) \qquad (17)$$

By replacing eq. (17) into eq. (16):

$$|\psi_0^{(2)}\rangle = \left[(1/\sqrt{2})(c_+|p_+^z\rangle + c_-|p_-^z\rangle) \otimes |x_+\rangle + (1/\sqrt{2})(c_+|p_+^z\rangle - c_-|p_-^z\rangle) \otimes |x_-\rangle\right] \otimes |p_0^x\rangle \qquad (18)$$

Since there is no collapse, the second interaction establishes the correlation between $|\psi_0^{(2)}\rangle$ and the eigenstates $|p_+^x\rangle$ and $|p_-^x\rangle$ of the second pointer $P^x$:

$$|\psi^{(2)}\rangle = (1/\sqrt{2})(c_+|p_+^z\rangle + c_-|p_-^z\rangle) \otimes |x_+\rangle \otimes |p_+^x\rangle + (1/\sqrt{2})(c_+|p_+^z\rangle - c_-|p_-^z\rangle) \otimes |x_-\rangle \otimes |p_-^x\rangle \qquad (19)$$

or, equivalently,

$$\begin{aligned}|\psi^{(2)}\rangle = &(c_+/\sqrt{2})|x_+\rangle \otimes |p_+^z\rangle \otimes |p_+^x\rangle + (c_-/\sqrt{2})|x_+\rangle \otimes |p_-^z\rangle \otimes |p_+^x\rangle + \\ &+ (c_+/\sqrt{2})|x_-\rangle \otimes |p_+^z\rangle \otimes |p_-^x\rangle - (c_-/\sqrt{2})|x_-\rangle \otimes |p_-^z\rangle \otimes |p_-^x\rangle\end{aligned} \qquad (20)$$

Now the conditional probabilities that link the results of the second measurement with those of the first measurement can be computed on the basis of the final non-collapsed state of the whole composite system. For instance, we can compute the probability that the second pointer acquires the value $p_+^x$ in the second measurement, given that the first pointer acquired the value $p_+^z$ in the first measurement, as follows:

$$pr\left(p_+^x / p_+^z\right) = \frac{pr\left(p_+^x \wedge p_+^z\right)}{pr\left(p_+^z\right)} \tag{21}$$

The function $pr\left(p_+^x \wedge p_+^z\right)$ is a legitimate probability because $P^x$ and $P^z$ are orthogonal components of the momentum of the measured particle and, then, $\left[P^x, P^z\right] = 0$ (see Laura & Vanni 2008): this is the essential step in this argument since the probability of a conjunction cannot be computed in the non-commuting case. The commutation of the pointers in consecutive measurements always holds because they belong to different measuring devices which, for this reason, are represented by different Hilbert spaces. So, from the first term of eq. (20),

$$pr\left(p_+^x \wedge p_+^z\right) = |c_+|^2 / 2 \tag{22}$$

In turn, $pr\left(p_+^z\right)$ can be computed from the first and the third terms of eq. (20):

$$pr\left(p_+^z\right) = \left(|c_+|^2 / 2\right) + \left(|c_+|^2 / 2\right) = |c_+|^2 \tag{23}$$

Therefore, the conditional probability is obtained:

$$pr\left(p_+^x / p_+^z\right) = \frac{pr\left(p_+^x \wedge p_+^z\right)}{pr\left(p_+^z\right)} = \frac{|c_+|^2 / 2}{|c_+|^2} = \frac{1}{2} \tag{24}$$

This value agrees with the well-known result obtained by means of the collapse hypothesis.

The case of consecutive measurements of the same observable can be obtained as a particular case of the general argument just presented. If in the second measurement the observable to be measured is again $S^z$, then the basis rotation of eq. (17) is not necessary, and the second interaction leads the whole system to the state

$$\left|\psi^{(2)}\right\rangle = c_+ \left|z_+\right\rangle \otimes \left|p_+^z\right\rangle \otimes \left|p_+^z\right\rangle + c_- \left|z_-\right\rangle \otimes \left|p_-^z\right\rangle \otimes \left|p_-^z\right\rangle \tag{25}$$

Then, now the conditional probabilities are

$$pr\left(p_+^z / p_+^z\right) = \frac{pr\left(p_+^z \wedge p_+^z\right)}{pr\left(p_+^z\right)} = \frac{|c_+|^2}{|c_+|^2} = 1 \quad \text{and} \quad pr\left(p_-^z / p_+^z\right) = \frac{pr\left(p_-^z \wedge p_+^z\right)}{pr\left(p_+^z\right)} = \frac{0}{|c_+|^2} = 0 \tag{26}$$

as expected. These results can be easily generalized for any different observables and for any number of consecutive measurements.

These results show that, by contrast to what the defender of the orthodox collapse interpretation believes, the experimental observed correlations between the outcomes of consecutive measurements can be perfectly explained with no need of the collapse hypothesis. Therefore, if certain no-collapse interpretations have difficulties to account for consecutive measurements, this fact is not due to their lack of the collapse hypothesis, but it is the consequence of endowing reduced states with a feature alien to them.

Summing up, reduced density operators are mere calculation tools for computing expectation values not only in the interacting case: when two systems do not interact but are entangled, the assumption that their quantum states are represented by operators obtained by partial traces leads to conceptual troubles. In other words, reduced states are not the legitimate quantum states of the subsystems of a composite system; in Schlosshauer terms: "*Since the two systems A and B are entangled and the total composite system is still described by the superposition, it follows from the standard rules of quantum mechanics that no individual definite state can be attributed to either one of the subsystems. Reduced density matrices of entangled subsystems therefore represent improper mixtures*" (Schlosshauer 2007, p. 48). In this sense Schlosshauer is completely clear: reduced states are not the quantum states of the subsystems of a whole quantum system, since no individual definite quantum state can be attributed to them.

## 5. *What do reduced states refer to?*

Even if one admits that the reduced state $\rho_1^r$ cannot properly be said to be the quantum state of $S_1$, it is difficult to deny that it supplies a certain description of the quantum system. As noticed above, the claim that reduced states are coarse-grained states has appeared in the literature on decoherence (see, *e.g.* Gell-Mann & Hartle 1993, Omnès 1994, Anastopoulos 2002). Nevertheless, in general the claim does not go beyond pointing out the operation of tracing over the degrees of freedom of the environment. Here we will show that the precise sense in which $\rho_1^r$ provides this description can be understood by means of a generalized conception of coarse-graining.

In its traditional classical form, a coarse-grained description is based on a partition of a phase space into discrete and disjoint cells: this mathematical procedure defines a projector $\Pi$ (see Mackey 1989). In other words, traditional coarse-graining amounts to a projection whose action is to cancel some components of the state vector $\rho$ corresponding to the fine-grained state: only certain components are retained as meaningful in the coarse-grained description $\rho_{cg} = \Pi\rho$. If this idea is generalized, coarse-graining can be conceived as an operation that cancels some components of a vector representing a state. From this generalized viewpoint, a partial trace is a particular case of coarse-graining, since it also cancels certain components of the density operator on which it is applied.

Let us recall the definition of reduced operator, $\langle O \rangle_\rho = \langle O_1 \rangle_{\rho_1^r}$, where $O = O_1 \otimes I_2$ (see eq. (5)). Although for dimensional reasons the reduced state $\rho_1^r$ cannot be expressed as a direct projection $\Pi\rho$ of the quantum state $\rho \in H \otimes H$, with $H = H_1 \otimes H_2$, the expectation value $\langle O_1 \rangle_{\rho_1^r}$ can also be expressed as the expectation value of $O = O_1 \otimes I_2$ in a coarse-grained state $\rho_{cg} \in H \otimes H$:

$$\langle O_1 \rangle_{\rho_1^r} = \langle O \rangle_{\rho_{cg}} \tag{27}$$

The density operator $\rho_{cg}$ represents a coarse-grained state because it can be obtained as $\rho_{cg} = \Pi\rho$, and the projector $\Pi$ executes the following operation:

$$\Pi\rho = \left(Tr_{(2)} \rho\right) \otimes \tilde{\delta}_2 = \rho_1^r \otimes \tilde{\delta}_2 \tag{28}$$

where $\tilde{\delta}_2 \in H_2 \otimes H_2$ is a normalized identity operator with coefficients $\tilde{\delta}_{2\alpha\beta} = \delta_{\alpha\beta} / \sum_\gamma \delta_{\gamma\gamma}$. The operator $\Pi$ is a projector since

$$\Pi\Pi\rho = \left(Tr_{(2)} \Pi\rho\right) \otimes \tilde{\delta}_2 = \left(Tr_{(2)} \left(\rho_1^r \otimes \tilde{\delta}_2\right)\right) \otimes \tilde{\delta}_2 = \rho_1^r \otimes \tilde{\delta}_2 = \Pi\rho \tag{29}$$

It is quite clear that $\rho_{cg} = \Pi\rho$, although belongs to $H \otimes H$, is not the quantum state of the composite system $S = S_1 + S_2$ represented by $H = H_1 \otimes H_2$: it is a coarse-grained state that disregards certain information of the quantum state of $S$. It is interesting to note that $\rho_{cg}$, although not the quantum state of $S$, is a coarse-grained state of the whole *closed system $S$* that may evolve *non-unitarily*. Therefore, it is not the case that non-unitary evolutions are proper of open quantum systems.

On the other hand, if we trace off the degrees of freedom of $S_2$, we recover the reduced state of $S_1$:

$$Tr_{(2)} \rho_{cg} = Tr_{(2)} \left(\rho_1^r \otimes \tilde{\delta}_2\right) = \rho_1^r \tag{30}$$

This means that the coarse-grained state $\rho_{cg}$, which "erases" the components corresponding to $S_2$, supplies the same information about the open system $S_1$ as the reduced state $\rho_1^r$, but now from the viewpoint of the composite system $S$ (see Castagnino & Fortin 2013). Therefore, the reduced density operator $\rho_1^r$ can also be conceived as a kind of coarse-grained state of $S$ that disregards certain degrees of freedom considered as irrelevant.

It is worth emphasizing that the arguments presented here, pointing toward the careful distinction between proper quantum states and reduced states, do not amount to be committed to "The Church of the Large Hilbert Space". This is the name coined by John Smolin to describe the habit of always conceiving a mixed state as the result of a partial trace over a larger system, and any non-unitary evolution as being embedded in some unitary evolution of a larger system. Certainly, our argumentation points to the fact that, as Paz and Zurek themselves admit, the fundamental dynamical law is the unitary time-evolution described by the dynamical postulate of the theory, and any other evolution must be defined in terms of that one. Nevertheless, this does not amount to consider pure states as the only proper quantum states of the theory. On the contrary, by following other authors (see, *e.g*. Ballentine 1998), in our arguments will have considered that the fundamental quantum state of the theory is represented by an operator that can denote both pure and mixed states. In other words, not all mixed states can be thought of as the result of a partial trace: there are mixed states that are legitimate quantum states not obtained by partial trace, which follow unitary evolutions and embody quantum correlations. The subtle point is to keep clear the difference between those quantum states and the reduced states, especially in interpretive contexts.[7]

Once the reduced state $\rho_1^r$ is viewed as a coarse-grained state, its non-unitary evolution does not restrict the application of the dynamical postulate to the quantum states of closed systems nor require a new dynamical postulate to be accounted for (recall the discussion of Subsection 2.b). The non-unitary evolution of $\rho_1^r$ turns out to be a situation analogous to the familiar case of classical instability, where it is completely natural to obtain a non-unitary coarse-grained evolution from the underlying unitary dynamics of the unstable system, with no need of restrictions or reformulations of the classical dynamical laws (see, *e.g*., Berkovitz, Frigg & Kronz 2006).[8] An author who has emphasized the analogy between the classical statistical case and the quantum case is Omnès (2001, 2002); he has repeatedly claimed that decoherence is a particular case of the phenomenon of irreversibility. Now the acknowledgement of the coarse-grained nature of the reduced state $\rho_S^r$ allows us to endow that claim with a more precise meaning. As in the case of classical instability, where the coarse-grained state $\rho_{cg}(t)$ approaches a final state $\rho_{cg(eq)}$ in spite of the unitary law ruling the evolution of the dynamical state, in decoherence the state $\rho_S^r(t)$ approaches a state $\rho_{S(dec)}^r$ ($\rho_S^r(t) \to \rho_{S(dec)}^r$), in spite of the fact that the quantum state indefinitely follows its unitary evolution.

---

[7] We are grateful to one of the referees for giving us the opportunity of emphasizing this point.

[8] Of course, this does not mean that the non-unitary evolution of the open subsystems of a quantum system is due to instability. The analogy emphasized here is based on the fact that both in unstable classical systems and in open quantum systems non-unitarily is obtained as the result of a coarse-graining on a underlying unitary evolution.

As it was said above, even if one accepted that a quantum mixed state could be interpreted in terms of the ignorance about the definite values of certain observables (an assumption that we will not discuss here), this interpretation is useless for decoherence because the reduced state $\rho_S^r$ of the decohering system is not its quantum mixed state. Now one can take a further step: when the coarse-grained origin of the reduced state is acknowledged, it is not difficult to see that decoherence is a relative phenomenon (for this claim, see Castagnino, Laura & Lombardi 2007). In fact, since many different coarse-grainings can be applied on a same closed system, there are many ways in which the whole closed system can be partitioned into a system of interest $S$ and its environment $E$, and the occurrence of decoherence or not must be studied for each particular partition (for the study of decoherence in different partitions of a single system, see Castagnino, Fortin & Lombardi 2010a, 2010b).

## 6. *Conclusions*

As stressed in the Introduction, reduced states obtained by partial traces are commonly considered as the quantum states of the subsystems of a closed system. Here we have argued against this assumption, both in the case that the subsystems interact with each other –the case of decoherence–, and in the case that there is no interaction between them –the case of no-collapse accounts of measurement–. Finally, we have shown that the reduced state of a subsystem has to be viewed as a coarse-grained state of the composite system to which it belongs. In summary, according to our view:

− Quantum states, pure or mixed, always follow unitary evolutions and embody quantum correlations.

− Given a composite system $S$ in a pure or mixed quantum state $\rho$, in the generic case –when the subsystems interact or are entangled with each other–:
  - no quantum state, pure or mixed, can be attributed to the subsystems $S_i$ of $S$.

- the reduced states $\rho_i^r$ of the subsystems $S_i$ of $S$
  * are not the quantum states of the $S_i$, because reduced states may evolve non unitarily and cancel correlations, and
  * can be conceived as a kind of coarse-grained states of the composite system $S$, which disregard certain degrees of freedom considered as irrelevant.

The argumentation of the present work must not be understood as a mere semantic discussion about the label to be attached to operators obtained by partial trace. Once we deprive reduced states of their role as quantum states of subsystems, new perspectives open up to our consideration. On the one hand, we can undertake a new reading of the phenomenon of decoherence, a reading that might account for situations not included in the orthodox environment-induced view (see Lombardi, Fortin & Castagnino 2012). On the other hand, we are led to rethink no-collapse interpretations, and the role played by reduced states in the selection of the definite-valued observables in quantum systems (see Ardenghi, Lombardi and Narvaja 2012). As a consequence, the physical meaning of reduced states deserves to be seriously discussed, since it has substantial consequences for the foundations of quantum mechanics.

## *Acknowledgements*

This work has been supported by grants of Consejo Nacional de Investigaciones Científicas y Técnicas (CONICET); Agencia Nacional de Promoción Científica y Tecnológica (ANPCyT); and Universidad de Buenos Aires (UBA), Argentina.

## *References*

Adler, S. L. (1995). *Quaternionic Quantum Mechanics and Quantum Fields*. New York: Oxford University Press.

Adler, S. (2003). "Why decoherence has not solved the measurement problem: A response to P. W. Anderson," *Studies in History and Philosophy of Modern Physics*, **34**: 135-142.

Anastopoulos, C. (2002). "Frequently asked questions about decoherence," *International Journal of Theoretical Physics*, **41**: 1573-1590.

Anderson, P. W. (2001). "Science: A 'dappled world' or a 'seamless web'?," *Studies in History and Philosophy of Modern Physics*, **34**: 487-494.

Ardenghi, J. S., Castagnino, M. and Lombardi, O. (2009). "Quantum mechanics: modal interpretation and Galilean transformations," *Foundations of Physics*, **39**: 1023-1045.

Ardenghi, J. S. and Lombardi, O. (2011). "The modal-Hamiltonian interpretation of quantum mechanics as a kind of "atomic" interpretation," *Physics Research International*, **2011**: 379604.

Ardenghi, J. S., Lombardi, O. and Narvaja, M. (2012). "Modal interpretations and consecutive measurements." Pp. 207-217, in V. Karakostas and D. Dieks (eds.), *EPSA 2011: Perspectives and Foundational Problems in Philosophy of Science*. Dordrecht: Springer.

Auletta, G. (2000). *Foundations and Interpretation of Quantum Mechanics*. Singapore: World Scientific.

Bacciagaluppi, G. (2008). "The role of decoherence in quantum mechanics." In E. N. Zalta (ed.), *The Stanford Encyclopedia of Philosophy* (Fall 2008 Edition), URL = <http://plato.stanford.edu/archives/fall2008/entries/qm-decoherence/>.

Ballentine, L. (1998). *Quantum Mechanics: A Modern Development*. Singapore: World Scientific.

Berkovitz, J., Frigg, R. and Kronz, F. (2006). "The ergodic hierarchy, randomness and Hamiltonian chaos," *Studies in History and Philosophy of Modern Physics*, **37**: 661-691.

Bricmont, J. (1995). "Science of chaos or chaos in science?," *Physicalia Magazine*, **17**: 159-208.

Bub, J. (1992). "Quantum mechanics without the projecton postulate," *Foundations of Physics*, **22**: 737-754.

Bub, J. (1997). *Interpreting the Quantum World*. Cambridge: Cambridge University Press.

Bub, J. and Clifton, R. (1996). "A uniqueness theorem for 'no collapse' interpretations of quantum mechanics," *Studies in History and Philosophy of Modern Physics*, **27**: 181-219.

Callender, C. (1999). "Reducing thermodynamics to statistical mechanics: the case of entropy," *Journal of Philosophy*, **96**: 348-373.

Cartwright, N. (1983). *How the Laws of Physics Lie*. Oxford: Clarendon Press.

Castagnino, M. and Fortin, S. (2013). "Formal features of a general theoretical framework for decoherence in open and closed systems," *International Journal of Theoretical Physics*, **52**: 1379-1398.

Castagnino, M., Fortin, S. and Lombardi, O. (2010a). "Suppression of decoherence in a generalization of the spin-bath model," *Journal of Physics A: Mathematical and Theoretical*, **43**: # 065304.

Castagnino, M., Fortin, S. and Lombardi, O. (2010b). "The effect of random coupling coefficients on decoherence," *Modern Physics Letters A*, **25**: 611-617.

Castagnino, M., Laura, R. and Lombardi, O. (2007). "A general conceptual framework for decoherence in closed and open systems," *Philosophy of Science*, **74**: 968-980.

Cohen-Tannoudji, C., Diu, B. and Lalöe, F. (1977). *Quantum Mechanics*. New York: John Wiley & Sons.

d'Espagnat, B. (1976). *Conceptual Foundations of Quantum Mechanics*. Reading MA: Benjamin.

d'Espagnat, B. (1995). *Veiled Reality. An Analysis of Present-Day Quantum Mechanical Concepts*. Reading MA: Addison-Wesley.

Dieks, D. (1988). "The formalism of quantum theory: an objective description of reality?," *Annalen der Physik*, **7**: 174-190.

Dieks, D. (1989). "Quantum mechanics without the projection postulate and its realistic interpretation," *Foundations of Physics*, **38**: 1397-1423.

Dieks, D. (1998). "Preferred factorizations and consistent property attribution." Pp. 144-160, in R. Healey and G. Hellman (eds.), *Quantum Measurement: Beyond Paradox*. Minneapolis: University of Minnesota Press.

Dieks, D. (1994a). "Objectification, measurement and classical limit according to the modal interpretation of quantum mechanics." Pp. 160-167, in P. Busch, P. Lahti and P. Mittelstaedt (eds.), *Proceedings of the Symposium on the Foundations of Modern Physics*. Singapore: World Scientific.

Dieks, D. (1994b). "Modal interpretation of quantum mechanics, measurements, and macroscopic behaviour," *Physical Review A*, **49**: 2290-2300.

Dieks, D. (2007). "Probability in modal interpretations of quantum mechanics," *Studies in History and Philosophy of Modern Physics*, **19**: 292-310.

Dieks, D. and Vermaas, P. E. (1998). *The Modal Interpretation of Quantum Mechanics*. Dordrecht: Kluwer Academic Publishers.

Driebe, D. J. (1994). "Letters (answer to Lebowitz, 1993)," *Physics Today*, **47**: 14-15.

Earman, J. and Rédei, M. (1996). "Why ergodic theory does not explain the success of equilibrium statistical mechanics," *British Journal for the Philosophy of Science*, **47**: 63-78.

Frigg, R. (2007). "A field guide to recent work on the foundations of thermodynamics and statistical mechanics." Pp. 99-196, in D. Rickles (ed.), *The Ashgate Companion to the New Philosophy of Physics*. London: Ashgate.

Gell-Mann, M. and Hartle, J. B. (1993). "Classical equations for quantum systems," *Physical Review D*, **47**: 3345-3382.

Harshman, N. L. and Wickramasekara, S. (2007a). "Galilean and dynamical invariance of entanglement in particle scattering," *Physical Review Letters*, **98**: 080406.

Harshman, N. L. and Wickramasekara, S. (2007b). "Tensor product structures, entanglement, and particle scattering," *Open Systems and Information Dynamics*, **14**: 341-351.

Healey, R. (1995). "Dissipating the quantum measurement problem," *Topoi*, **14**: 55-65.

Healey, R. (2013). "Physical composition," *Studies in History and Philosophy of Modern Physics*, **44**: 48-62.

Hughes, R. I. G. (1989). *The Structure and Interpretation of Quantum Mechanics*. Cambridge MA: Harvard University Press.

Joos, E. (2000). "Elements of environmental decoherence." Pp. 1-17, in P. Blanchard, D. Giulini, E. Joos, C. Kiefer and I.-O. Stamatescu (eds.), *Decoherence: Theoretical, Experimental, and Conceptual Problems, Lecture Notes in Physics, Vol. 538*. Heidelberg-Berlin: Springer.

Kochen, S. (1985). "A new interpretation of quantum mechanics." Pp. 151-169, in P. Mittelstaedt and P. Lahti (eds.), *Symposium on the Foundations of Modern Physics 1985*. Singapore: World Scientific.

Koopman, B. O. (1931). "Hamiltonian systems and transformations in Hilbert space," *Proceedings of the National Academy of Sciences*, **18**: 315-318.

Laura, R. and Vanni, L. (2008). "Conditional probabilities and collapse in quantum measurements," *International Journal of Theoretical Physics*, **47**: 2382-2392.

Lebowitz, J. L. (1993). "Boltzmann's entropy and time's arrow," *Physics Today*, **46**: 32-38.

Lebowitz, J. L. (1994a). "Lebowitz replies," *Physics Today*, **47**: 115-116.

Lebowitz, J. L. (1994b). "Statistical mechanics: a selective review of two central issues," *Reviews of Modern Physics*, **71**: S346-S357.

Leggett, A. J. (1987). "Reflections on the quantum measurement paradox." Pp. 85-104, in B. J. Hiley and F. D. Peat (eds.), *Quantum Implications*. London: Routledge and Kegan Paul.

Lombardi, O. (2003). "El problema de la ergodicidad en mecánica estadística," *Crítica. Revista Hispanoamericana de Filosofía*, **35**: 3-41.

Lombardi, O. and Castagnino, M. (2008). "A modal-Hamiltonian interpretation of quantum mechanics," *Studies in History and Philosophy of Modern Physics*, **39**: 380-443.

Lombardi, O., Castagnino, M. and Ardenghi, J. S. (2010). "The modal-Hamiltonian interpretation and the Galilean covariance of quantum mechanics," *Studies in History and Philosophy of Modern Physics*, **41**: 93-103.

Lombardi, O. and Dieks, D. (2012). "Modal interpretations of quantum mechanics." In E. N. Zalta (ed.), *The Stanford Encyclopedia of Philosophy* (Winter 2012 Edition), URL = <http://plato.stanford.edu/archives/win2012/entries/qm-modal/>.

Lombardi, O., Fortin, S. and Castagnino, M. (2012). "The problem of identifying the system and the environment in the phenomenon of decoherence." Pp. 161-174, in H. de Regt, S. Okasha and S. Hartmann (eds.), *EPSA Philosophy of Science: Amsterdam 2009*. Dordrecht: Springer.

Mackey, M. C. (1989). "The dynamic origin of increasing entropy," *Review of Modern Physics*, **61**: 981-1015.

Masillo, F., Scolarici, G. & Sozzo, S. (2009). "Proper versus improper mixtures: towards a quaternionic quantum mechanics," *Theoretical and Mathematical Physics*, **160**: 1006-1013.

Messiah, A. (1961). *Quantum Mechanics*, Vol. 1. Amsterdam: North-Holland.

Misra. B., Prigogine, I. and Courbage, M. (1979). "From deterministic dynamics to probabilistic descriptions," *Physica A*, **98**: 1-26.

Nicolis, G. and Prigogine, I. (1989). *Exploring Complexity. An Introduction*. New York: Freeman & Company.

Ney, A. and Albert, D. (2013). *The Wave Function*. New York: Oxford University Press.

Omnès, R. (1994). *The Interpretation of Quantum Mechanics*. Princeton: Princeton University Press.

Omnès, R. (2001). "Decoherence: An irreversible process," *Los Alamos National Laboratory*, arXiv:quant-ph/0106006.

Omnès, R. (2002). "Decoherence, irreversibility and the selection by decoherence of quantum states with definite probabilities," *Physical Review A*, **65**: 052119.

Pauli, W. (1933). "Die allgemeinen Prinzipien der Wellenmechanik." Pp. 83-272, in H. Geiger and K. Scheel (eds.), *Handbuch der Physik, Vol. 24*. Berlin: Springer. English translation: *General Principles of Quantum Mechanics*, Berlin: Springer, 1980.

Paz, J. P. and Zurek, W. H. (2002). "Environment-induced decoherence and the transition from quantum to classical." Pp. 77-148, in D. Heiss (ed.), *Fundamentals of Quantum Information, Lecture Notes in Physics, Vol. 587*. Heidelberg-Berlin: Springer (the page numbers are taken from arXiv:quant-ph/0010011).

Schlosshauer, M. (2007). *Decoherence and the Quantum-to-Classical Transition*. Heidelberg-Berlin: Springer (4th reprint 2009).

Uffink, J. (2007). "Compendium of the foundations of classical statistical physics," Pp. 923-1074, in J. Butterfield and J. Earman (eds.), *Philosophy of Physics*. Amsterdam: Elsevier.

Vermaas, P. and Dieks, D. (1995). "The modal interpretation of quantum mechanics and its generalization to density operators," *Foundations of Physics*, **25**: 145-158.

Zeh, D. (1970). "On the interpretation of measurement in quantum theory," *Foundations of Physics*, **1**: 69-76.

Zeh, D. (1973). "Toward a quantum theory of observation," *Foundations of Physics*, **3**: 109-116.

Zeh, H. D. (2005). "Roots and fruits of decoherence," *Séminaire Poincaré*, **2**: 1-19.

Zurek, W. H. (1981). "Pointer basis of quantum apparatus: into what mixtures does the wave packet collapse?," *Physical Review D*, **24**: 1516-1525.

Zurek, W. H. (1982). "Environment-induced superselection rules," *Physical Review D*, **26**: 1862-1880.

Zurek, W. H. (1991). "Decoherence and the transition from quantum to classical," *Physics Today*, **44**: 36-44.

Zurek, W. H. (1993). "Preferred states, predictability, classicality and the environment-induced decoherence," *Progress of Theoretical Physics*, **89**: 281-312.

Zurek, W. H. (1994). "Preferred sets of states, predictability, classicality and environment-induced decoherence." Pp. 175-207, in J. J. Halliwell, J. Pérez-Mercader and W. H. Zurek (eds.), *Physical Origins of Time Asymmetry*. Cambridge: Cambridge University Press.

Zurek, W. H. (2003). "Decoherence, einselection, and the quantum origins of the classical," *Reviews of Modern Physics*, **75**: 715-776.